\documentclass[groupedaddress, superscriptaddress, citesort]{revtex4}
\usepackage{amssymb, amsmath}
\usepackage[dvips]{graphicx}

\def\be{\begin{equation}}
\def\ee{\end{equation}}
\def\ba{\begin{array}}
\def\ea{\end{array}}

\begin{document}
\baselineskip=18pt

\title {Dynamics of quantum battery capacity under Markovian channels}
\author{Yao-Kun Wang}
\affiliation{College of Mathematics,  Tonghua Normal University, Tonghua, Jilin 134001, China}
\author{Li-Zhu Ge}
\affiliation{The Branch Campus of Tonghua Normal University, Tonghua, Jilin 134001, China}
\author{Tinggui Zhang}
\affiliation{School of Mathematics and Statistics,  Hainan Normal University, Haikou, 571158, China}
\author{Shao-Ming Fei}
\affiliation{School of Mathematical Sciences,  Capital Normal
University,  Beijing 100048,  China}
\author{Zhi-Xi Wang}
\affiliation{School of Mathematical Sciences,  Capital Normal
University,  Beijing 100048,  China}

\begin{abstract}
We study the dynamics of the quantum battery capacity for the Bell-diagonal states under Markovian channels on the first subsystem. We show that the capacity increases for special Bell-diagonal states under amplitude damping channel. The sudden death of the capacity occurs under depolarizing channel. We also investigate the capacity evolution of Bell-diagonal states under Markovian channels on the first subsystem $n$ times. It is shown that the capacity under depolarizing channel decreases initially, then increases for small $n$ and tend to zero for large $n$. We find that under bit flip channel and amplitude damping channel, the quantum battery capacity of special Bell-diagonal states tends to a constant for large $n$, namely, the frozen capacity occurs. The dynamics of the capacity of the Bell-diagonal states under two independent same type local Markovian channels is also studied.
\end{abstract}

\maketitle

\section{Introduction}
With the rapid development of quantum information and quantum thermodynamics, some novel quantum devices have been introduced recently. The quantum battery presented by R. Alicki and M. Fannes shows the maximum amount of energy that can be extracted from a quantum system in an informational theoretic view\cite{ramf}. Since then many interesting results have been obtained \cite{mkmh,mkmp,egnm,gjfm,fffl,cdjo,tjkm,dmgv,gfrz,gdav,rali,gmav,dgmp,agnj,dgdm,sfff,stas,ammd,khgw,sfss,yyxs,srvg,ksus,ffqz,stsa,mtjj,fyja,fyyj,hsqx,tlsa,khzh,rbgp,fhja,srsv,jcrj, fmaj,sgas,dmam,fsjm,paku,asov,bham,tscg,ksus2,fdfy,srsv2,xymr,jsjj,khhg,rgdv,pbfn,hhqk}, including the construction of charging models \cite{fffl,gjfm,dmgv,gfrz,gmav,fmaj,jcrj,hsqx} and many-body quantum batteries\cite{tjkm,dgmp,yyxs,sfss,khgw,fyyj,fyja,asov}. Among them, Ferraro et al. proved that the quantum battery charges faster than the ordinary batteries in collective charging mode\cite{dmgv}. A series of other quantum battery models have been also brought up\cite{tjkm,dgdm,hhqk}, together with some experimental studies\cite{jjts,aprm,jkld,jjis,ndfa}, see \cite{fsjm} for a review of quantum batteries.

An important indicator of quantum batteries is the battery capacity\cite{srsv,xymr}. Recently, the authors in Ref. \cite{xymr} provided a new definition of quantum battery capacity,
\be\label{zyf1}
C(\rho,H):=\sum_{i=0}^{d-1}\epsilon_i(\lambda_i-\lambda_{d-1-i}),
\ee
where $\lambda_0\leq\lambda_1\leq\cdots \leq\lambda_{d-1}$ represent the eigenvalues of the quantum state $\rho$ and $\epsilon_0\leq\epsilon_1\leq\cdots \leq\epsilon_{d-1}$ the
eigenenergies of the Hamiltonian $H=\sum_i\epsilon_i|\varepsilon_i\rangle\langle\varepsilon_i|$. $C(\rho,H)$ is a Schur-convex functional of the quantum state
and does not alter whenever the battery is unitarily charged.

In this article, we investigate the evolution of quantum battery capacity in Bell-diagonal states, when the first subsystem undergoes Markovian channels. We show that the quantum battery capacity of the Bell-diagonal states increases under amplitude damping channel. In particular, we find that the phenomenon of sudden death of the quantum battery capacity occurs for special Bell-diagonal states under depolarizing channel. We also study the capacity evolution of Bell-diagonal states when the first subsystem undergoes $n$ times Markovian channels. It is shown that the capacity under depolarizing channel decreases initially, and then increases for small $n$ and tends to $0$ for large $n$. We also find that phenomenon of frozen of capacity occurs for special Bell-diagonal states under bit flip channel and amplitude damping channel for large $n$. In addition, we investigate the dynamics of capacity of the Bell-diagonal states under two independent local Markovian channels of the same type.

\section{Quantum battery capacity of Bell-diagonal states under Markovian channels}

We consider the initial states to be the two-qubit Bell-diagonal ones,
\begin{equation}
\rho_{AB}=\frac{1}{4}(I_2\otimes I_2+\sum_{i=1}^3c_i\sigma_i\otimes \sigma_i),\label{bs}
\end{equation}
where $\sigma_1,\sigma_2,\sigma_3$ are the standard
Pauli matrices, $c_i$, $i=1,2,3$, are real constants such that $\rho_{AB}$ is a well defined density matrix. $\rho_{AB}$ has eigenvalues
$\lambda_0=(1-c_1-c_2-c_3)/4$, $\lambda_1=(1-c_1+c_2+c_3)/4$, $\lambda_2=(1+c_1-c_2+c_3)/4$ and $\lambda_3=(1+c_1+c_2-c_3)/4$. The coefficients $c_i$ are chosen such that
$\lambda_j\in[0,1]$ for $j=0,1,2,3$.

Let us consider the following Hamiltonian for the two-qubit system, $H_{AB}=\epsilon^A\sigma_3\otimes I_2+\epsilon^B I_2\otimes \sigma_3$,
where $\epsilon^A \geq \epsilon^B \geq 0$. The eigenvalues of $H_{AB}$ are $-\epsilon^A-\epsilon^B$, $-\epsilon^A+\epsilon^B$,
$\epsilon^A-\epsilon^B$ and $\epsilon^A+\epsilon^B$ in ascending order.
Therefore, if $c_{1}>c_{2}>c_{3}>0$, then $\lambda_0< \lambda_1< \lambda_2< \lambda_3$. From (\ref{zyf1}) we have
\begin{equation}
C_{123}(\rho_{AB},H_{AB})=(c_1+c_2)(\epsilon^A+\epsilon^B)+(c_1-c_2)(\epsilon^A-\epsilon^B).\label{123}
\end{equation}
For $c_{1}>c_{3}>c_{2}>0$, we have $\lambda_0< \lambda_1< \lambda_3< \lambda_2$ and
\begin{equation}
C_{132}(\rho_{AB},H_{AB})=(c_1+c_3)(\epsilon^A+\epsilon^B)+(c_1-c_3)(\epsilon^A-\epsilon^B).\label{132}\nonumber
\end{equation}
For $c_{2}>c_{1}>c_{3}>0$, we have $\lambda_0< \lambda_2< \lambda_1< \lambda_3$ and
\begin{equation}
C_{213}(\rho_{AB},H_{AB})=(c_2+c_1)(\epsilon^A+\epsilon^B)+(c_2-c_1)(\epsilon^A-\epsilon^B).\label{213}\nonumber \end{equation}
For $c_{2}>c_{3}>c_{1}>0$, we have $\lambda_0< \lambda_2< \lambda_3< \lambda_1$ and
\begin{equation}
C_{231}(\rho_{AB},H_{AB})=(c_2+c_3)(\epsilon^A+\epsilon^B)+(c_2-c_3)(\epsilon^A-\epsilon^B).\label{231}
\end{equation}
For $c_{3}>c_{1}>c_{2}>0$, we have $\lambda_0< \lambda_3< \lambda_1< \lambda_2$ and
\begin{equation}
C_{312}(\rho_{AB},H_{AB})=(c_3+c_1)(\epsilon^A+\epsilon^B)+(c_3-c_1)(\epsilon^A-\epsilon^B).\label{312}\nonumber
\end{equation}
For $c_{3}>c_{2}>c_{1}>0$, we have $\lambda_0< \lambda_3< \lambda_2< \lambda_1$ and
\begin{equation}
C_{321}(\rho_{AB},H_{AB})=(c_3+c_2)(\epsilon^A+\epsilon^B)+(c_3-c_2)(\epsilon^A-\epsilon^B).\label{321}\nonumber
\end{equation}
The reduced states of $\rho_{AB}$ are $\rho_A=I_2/2$ and $\rho_B=I_2/2$, respectively. Consequently, we have $C(\rho_A,H_A)=0$.

(1) \emph{Quantum battery capacity evolution of Bell-diagonal states under Markovian channels on the first subsystem}

We will focus on the evolution of a quantum state
$\rho$ under a trace-preserving quantum operation $\varepsilon(\rho)$\cite{nielsen},
$\varepsilon(\rho) = \sum_{i,j} \left(E_i\otimes E_j\right) \rho \left(E_i \otimes E_j\right)^\dagger$, where $\{E_k\}$ is the set of Kraus operators with $\sum_k E_k^\dagger E_k = I$. The Kraus operators of some typical channels are listed by in Table~\ref{t1} \cite{wang}, where the decoherence processes bit flip channel (bf), phase flip channel (pf),
bit-phase flip channel(bpf) and depolarizing channel (dep) preserve the Bell-diagonal form of the density operator $\rho$. For the case of generalized amplitude damping channel (gad),
the Bell-diagonal form is kept for arbitrary $\gamma$ and $p=1/2$. In this situation, one has
\begin{equation}
\rho^\prime=\varepsilon(\rho)=\frac{1}{4}(I\otimes I+\sum_{i=1}^3c^\prime_i\sigma_i\otimes\sigma_i)\label{newstate},
\end{equation}
where $c_{1}^\prime, c_{2}^\prime, c_{3}^\prime\in[-1,1]$ are given in Table~\ref{t2} \cite{wang}.
\begin{table}
\begin{center}
\begin{tabular}{c c}
\hline \hline
 & $\textrm{Kraus operators}$                                         \\  \hline & \\
bf   & $E_0 = \sqrt{1-p/2}\, I ,~~~ E_1 = \sqrt{p/2} \,\sigma_1$                        \\
 & \\
pf   & $E_0 = \sqrt{1-p/2}\, I ,~~~ E_1 = \sqrt{p/2}\, \sigma_3$                        \\
 & \\
bpf & $E_0 = \sqrt{1-p/2}\, I ,~~~ E_1 = \sqrt{p/2} \,\sigma_2$                        \\
 & \\
dep & $E_0 = \sqrt{1-p}\, I ,~~~ E_1 = \sqrt{p/3} \,\sigma_1$                        \\
 & \\
 & $E_2 = \sqrt{p/3}\,\sigma_2 ,~~~ E_3= \sqrt{p/3} \,\sigma_3$                        \\
 & \\
gad   &
$E_0=\sqrt{p}\left(
\begin{array}{cc}
1 & 0 \\
0 & \sqrt{1-\gamma} \\
\end{array} \right) ,~~~~
E_2=\sqrt{1-p}\left(
\begin{array}{cc}
\sqrt{1-\gamma} & 0 \\
0 & 1 \\
\end{array} \right)$  \\
& \\
 & $E_1=\sqrt{p}\left(
\begin{array}{cc}
0 & \sqrt{\gamma} \\
0 & 0 \\
\end{array} \right) ,~~~~
E_3=\sqrt{1-p}\left(
\begin{array}{cc}
0 & 0 \\
\sqrt{\gamma} & 0 \\
\end{array} \right)$  \\ \hline \hline
\end{tabular}
\caption[table 1]{Kraus operators for the qubit quantum channels: bit flip channel (bf), phase flip channel (pf),
bit-phase flip channel(bpf), depolarizing channel (dep), and generalized amplitude damping channel (gad), where
$p$ and $\gamma$ are decoherence probabilities, $0<p<1$, $0<\gamma<1$.}
\label{t1}
\end{center}
\end{table}

When $c_{1}>c_{2}>c_{3}>0$, from Table~\ref{t2}, $c^\prime_1, c^\prime_2, c^\prime_3$ of $\rho^\prime$ under bit flip channel (bf), depolarizing channel (dep), and generalized amplitude damping channel (gad) satisfy the
condition for Eq. (\ref{123}), i.e., $c^\prime_1> c^\prime_2> c^\prime_3$. Let $C_{123}^{bf}(\rho^\prime,H_{AB})$,  $C_{123}^{dep}(\rho^\prime,H_{AB})$ and $C_{123}^{gad}(\rho^\prime,H_{AB})$ be quantum battery capacity of $\rho^\prime$ under bit flip channel, depolarizing channel and generalized amplitude damping channel, respectively. By Eq. (\ref{newstate}) and (\ref{123}), we have
\begin{equation}\nonumber
\begin{array}{rcl}
C_{123}^{bf}(\rho^\prime,H_{AB})&=&(c_1+c_2 (1-p)^{2})(\epsilon^A+\epsilon^B)+(c_1-c_2 (1-p)^{2})(\epsilon^A-\epsilon^B),\\[1.5mm]
C_{123}^{dep}(\rho^\prime,H_{AB})&=&(c_1+c_2)(1-4p/3)
(\epsilon^A+\epsilon^B)+(c_1-c_2)(1-4p/3)(\epsilon^A-\epsilon^B),\\[1.5mm]
C_{123}^{gad}(\rho^\prime,H_{AB})&=&(c_1+c_2)(1-p)(\epsilon^A+\epsilon^B)
+(c_1-c_2)(1-p)(\epsilon^A-\epsilon^B).
\end{array}
\end{equation}

\begin{table}
\begin{center}
\begin{tabular}{c c c c}
\hline \hline
$\textrm{Channel}$ & $c^\prime_1$      & $c^\prime_2$     & $c^\prime_3$      \\ \hline
& & & \\
bf                 &  $c_1$            & $c_2 (1-p)^2$    & $c_3 (1-p)^2$     \\
& & & \\
pf                 &  $c_1 (1-p)^2$    & $c_2 (1-p)^2$    & $c_3$             \\
& & & \\
bpf                &  $c_1 (1-p)^2$    & $c_2$            & $c_3 (1-p)^2$     \\
& & & \\
dep                &  $c_1 (1-4p/3)$   ~~~~~ & $c_2 (1-4p/3)$ ~~~~~       & $c_3 (1-4p/3)$     \\
& & & \\
gad                &  $c_1 (1-p)$ & $c_2 (1-p)$ & $c_3 (1-p)^2$ \\ \hline \hline
\end{tabular}
\caption[table 2]{Correlation coefficients for the quantum operations: bit flip channel (bf), phase flip channel (pf),
bit-phase flip channel (bpf), depolarizing channel (dep), and generalized amplitude damping channel (gad). For gad, we
have fixed $p=1/2$ and replaced $\gamma$ by $p$.}
\label{t2}
\end{center}
\end{table}

Under the amplitude damping channel, the Bell-diagonal states are transformed into the output state,
\begin{eqnarray}\label{sta1}
\rho_{adc}^\prime = \frac{1}{4} \left(
\begin{array}{cccc}
2-(1-c_3)(1-p) & 0 & 0 & (c_1 -c_2)(1-p)^{\frac{1}{2}} \\
0 & 2-(1+c_3)(1-p) & (c_1 +c_2)(1-p)^{\frac{1}{2}} & 0 \\
0 & (c_1 +c_2)(1-p)^{\frac{1}{2}} & (1-c_3)(1-p) & 0 \\
(c_1 -c_2)(1-p)^{\frac{1}{2}} & 0 & 0 & (1+c_3)(1-p)
\end{array}
\right).\nonumber
\end{eqnarray}
The eigenvalues of $\rho_{adc}^\prime$ are $u_{0}=1/4(1-c_{3}(1-p)-\sqrt{(c_{1}+c_{2})^{2}(1-p)+p^{2}})$, $u_{1}=1/4(1-c_{3}(1-p)+\sqrt{(c_{1}+c_{2})^{2}(1-p)+p^{2}})$,
$u_{2}=1/4(1+c_{3}(1-p)-\sqrt{(c_{1}-c_{2})^{2}(1-p)+p^{2}})$ and $u_{3}=1/4(1+c_{3}(1-p)+\sqrt{(c_{1}-c_{2})^{2}(1-p)+p^{2}})$. Set $c_{1}=0.5, c_{2}=0.3, c_{3}=0.1, \epsilon^{A}=0.6$ and $ \epsilon^{B}=0.3$. We have $u_{0}<u_{2}<u_{3}<u_{1}$,
see Fig. 1 (a). Using Eq. (\ref{zyf1}) we obtain
\begin{align}
C^{adc}_{123}(\rho^\prime_{adc},H_{AB})&=\sqrt{(c_{1}+c_{2})^{2}
(1-p)+p^{2}}(\epsilon^A+\epsilon^B)+\sqrt{(c_{1}-c_{2})^{2}(1-p)
+p^{2}}(\epsilon^A-\epsilon^B)\nonumber\\
&=0.9\sqrt{0.64(1-p)+p^{2}}+0.3\sqrt{0.04(1-p)+p^{2}}.\nonumber
\end{align}

From Fig. 1 (b), we see that for the initial Bell-diagonal states the quantum battery capacity decreases as $p$ increases under bit flip channel, depolarizing channel and generalized amplitude damping channel. Under generalized amplitude damping channel the quantum battery capacity approaches to zero as $p$ increases to $1$. Under depolarizing
channel the quantum battery capacity becomes zero as $p \geq 0.75$, a phenomenon of sudden death of quantum battery capacity. On the other hand, the quantum battery capacity increases as $p$ increases under amplitude damping channel.
\begin{figure}[h]
\begin{center}
\scalebox{1.0}{(a)}{\includegraphics[width=7cm]{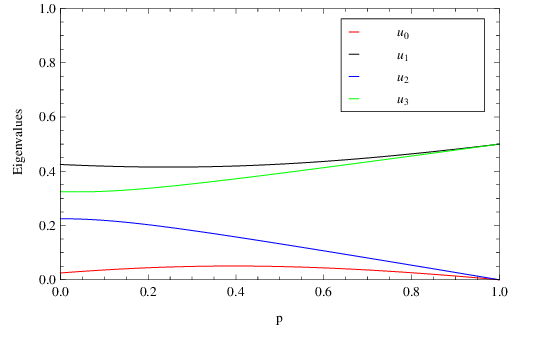}}
\scalebox{1.0}{(b)}{\includegraphics[width=7cm]{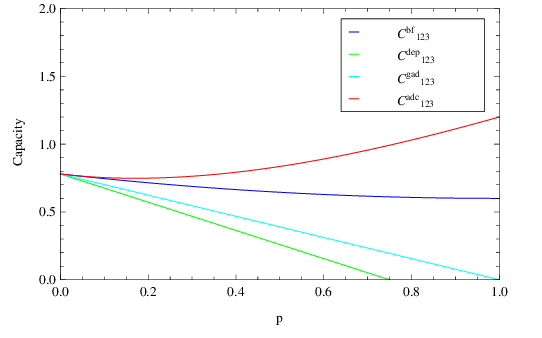}}
\caption{(a) The eigenvalues of  $\rho_{adc}^\prime$ as a function of $p$ for $c_{1}=0.5, c_{2}=0.3, c_{3}=0.1, \epsilon^{A}=0.6$ and $\epsilon^{B}=0.3 $. (b) Quantum battery capacity evolution for Bell-diagonal states with \{$c_{1}=0.5, c_{2}=0.3, c_{3}=0.1, \epsilon^{A}=0.6, \epsilon^{B}=0.3 $\} under bit flip channel ($C^{bf}_{123}$), depolarizing channel ($C^{dep}_{123}$), amplitude damping channel ($C^{adc}_{123}$) and generalized amplitude damping channel ($C^{gad}_{123}$) as a function of $p$.}
\end{center}
\end{figure}

When $c_{2}>c_{3}>c_{1}>0$, we have $c^\prime_2> c^\prime_3> c^\prime_1$ under bit-phase flip (bpf), depolarizing channel (dep), see Table~\ref{t2}. Let $C_{231}^{bpf}(\rho^\prime,H_{AB})$ and $C_{231}^{dep}(\rho^\prime,H_{AB})$ denote the quantum battery capacity of $\rho$ under bit-phase flip and depolarizing channel, respectively. By Eq. (\ref{newstate}) and (\ref{231}) we have
\begin{align}\nonumber
C_{231}^{bpf}(\rho^\prime,H_{AB})&=(c_2+c_3 (1-p)^{2})(\epsilon^A+\epsilon^B)+(c_2-c_3 (1-p)^{2})(\epsilon^A-\epsilon^B),\\ \nonumber
C_{231}^{dep}(\rho^\prime,H_{AB})&=(c_2+c_3)(1-4p/3)
(\epsilon^A+\epsilon^B)+(c_2-c_3)(1-4p/3)(\epsilon^A-\epsilon^B).\nonumber
\end{align}

Set $c_{1}=0.1, c_{2}=0.5, c_{3}=0.3, \epsilon^{A}=0.6$ and $\epsilon^{B}=0.3$. We have $u_{0}<u_{2}<u_{1}<u_{3}$, see Fig. 2. (a). Using Eq. (\ref{zyf1}) we have
\begin{align}
C^{adc}_{231}(\rho^\prime_{adc},H_{AB})&=
1/2(2c_{3}+\sqrt{(c_{1}-c_{2})^{2}(1-p)+p^{2}}+\sqrt{(c_{1}+c_{2})^{2}(1-p)+p^{2}})
(\epsilon^A+\epsilon^B)\nonumber\\
&~~~+1/2(-2c_{3}+\sqrt{(c_{1}+c_{2})^{2}(1-p)+p^{2}}+\sqrt{(c_{1}-c_{2})^{2}(1-p)
+p^{2}})(\epsilon^A-\epsilon^B)\nonumber\\
&=0.9/2(0.6(1-p)+\sqrt{(0.16(1-p)+p^{2}}+\sqrt{(0.36(1-p)+p^{2}})\nonumber\\
&~~~+0.3/2(-0.6(1-p)+\sqrt{(0.36(1-p)+p^{2}}+\sqrt{(0.16(1-p)+p^{2}}).\nonumber
\end{align}
The quantum battery capacity evolutions under bit-phase flip channel, depolarizing channel, generalized amplitude damping channel and amplitude damping channel are similar with the situation of $c_{1}>c_{2}>c_{3}>0$, see Fig. 2 (b).
\begin{figure}[h]
\begin{center}
\scalebox{1.0}{(a)}{\includegraphics[width=7cm]{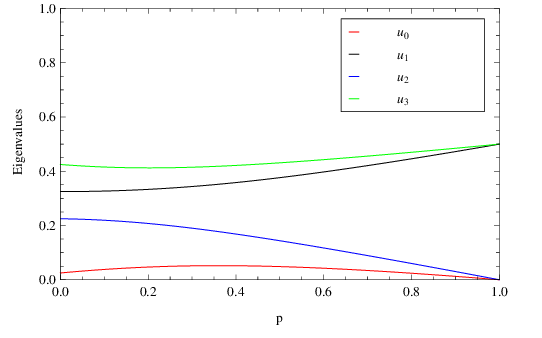}}
\scalebox{1.0}{(b)}{\includegraphics[width=7cm]{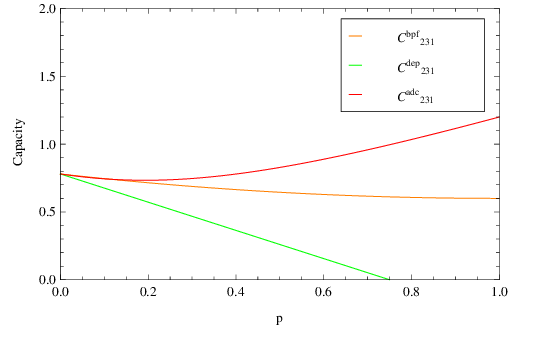}}
\caption{(a) The eigenvalues of $\rho_{adc}^\prime$ as a function of $p$ for $c_{1}=0.1, c_{2}=0.5, c_{3}=0.3, \epsilon^{A}=0.6, \epsilon^{B}=0.3$. (b) Quantum battery capacity evolution for Bell-diagonal states with \{$c_{1}=0.1, c_{2}=0.5, c_{3}=0.3, \epsilon^{A}=0.6, \epsilon^{B}=0.3 $\} under bit-phase flip channel ($C^{bpf}_{231}$), depolarizing channel ($C^{dep}_{231}$) and amplitude damping channel ($C^{adc}_{231}$)) as a function of $p$.}
\end{center}
\end{figure}

When $c_{3}>c_{2}>c_{1}>0$, the quantum battery capacity for the same Bell-diagonal state under phase flip channel is similar with the situations of bit flip channel {$c_{1}>c_{2}>c_{3}>0$} and bit-phase flip channel {$c_{2}>c_{3}>c_{1}>0$}.

\bigskip

(2) \emph{Quantum battery capacity evolution under $n$ times Markovian channels on the first subsystem}

\bigskip
Next, we consider the quantum battery capacity dynamics of Bell-diagonal states under $n$ times Markovian channels on the first subsystem. As the decoherence processes bf, pf, bpf, dep and gad preserve the Bell-diagonal form of the density operator, the parameters of the output state are given by $c^\prime_1$, $c^\prime_2$, $c^\prime_3$ in table \ref{t3} when a Bell-diagonal state goes through the channel $n$ times.
\begin{table}
\begin{center}
\begin{tabular}{c c c c}
\hline \hline
$\textrm{Channel}$ & $c^\prime_1$      & $c^\prime_2$     & $c^\prime_3$      \\ \hline
& & & \\
$bf^{n}$                 &  $c_1$            & $c_2 (1-p)^{2n}$    & $c_3 (1-p)^{2n}$     \\
& & & \\
$pf^{n}$                 &  $c_1 (1-p)^{2n}$    & $c_2 (1-p)^{2n}$    & $c_3$             \\
& & & \\
$bpf^{n}$                &  $c_1 (1-p)^{2n}$    & $c_2$            & $c_3 (1-p)^{2n}$     \\
& & & \\
$dep^{n}$  ~              &  $c_1 (1-4p/3)^{n}$ ~~~   & $c_2 (1-4p/3)^{n}$ ~~~       & $c_3 (1-4p/3)^{n}$     \\
& & & \\
$gad^{n}$                &  $c_1 (1-p)^{n}$ & $c_2 (1-p)^{n}$ & $c_3 (1-p)^{2n}$ \\ \hline \hline
\end{tabular}
\caption[table 3]{Correlation coefficients for $n$ times quantum channels: bit flip channel ($bf^{n}$), phase flip channel ($pf^{n}$), bit-phase flip channel ($bpf^{n}$), depolarizing channel ($dep^{n}$), and generalized amplitude damping channel($gad^{n}$). For the gad, we have fixed $p=1/2$ and replaced $\gamma$ by $p$.}
\label{t3}
\end{center}
\end{table}

Set $c_{1}=0.5, c_{2}=0.3, c_{3}=0.1,\epsilon^{A}=0.6$ and $\epsilon^{B}=0.3$. Using Eq. (\ref{123}), we obtain quantum battery capacity for Bell-diagonal states under $n$ times various Markovian noise channels, see Fig. 3. One sees that when $p$
approaches to $1$, the quantum battery capacity approaches to a constant for large $n$, that is, the frozen capacity appears, see Fig. 3 (a). When $p$ increases, the quantum battery capacity under depolarizing channel decreases initially and then increases. It is worth mentioning that the capacity tends to $0$ as $p$ approaches to $1$ for large $n$, see Fig. 3 (b). The quantum battery capacity under $n$ times generalized amplitude damping channel decreases as $p$ increases. The curvatures of the cures gradually become large for large $n$, see Fig. 3 (c).
\begin{figure}[h]
\begin{center}
\scalebox{1.0}{(a)}{\includegraphics[width=4.8cm]{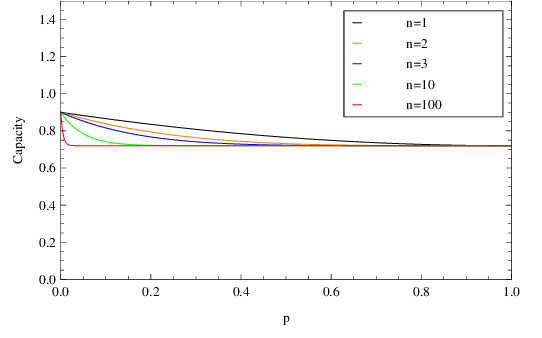}}
\scalebox{1.0}{(b)}{\includegraphics[width=4.8cm]{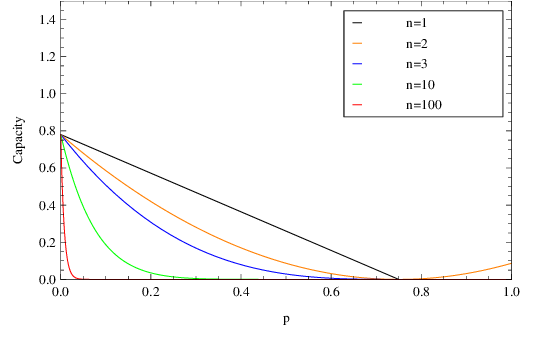}}
\scalebox{1.0}{(c)}{\includegraphics[width=4.8cm]{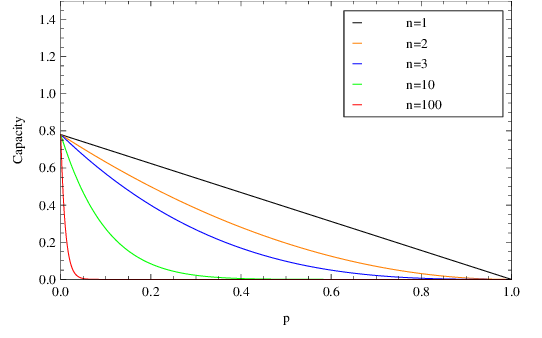}}
\caption{Quantum battery capacity for Bell-diagonal state \{$c_{1}=0.5$, $c_{2}=0.3$, $c_{3}=0.1$, $\epsilon^{A}=0.6, \epsilon^{B}=0.3$\}: (a) under bit flip channel $n$ times, (b) under depolarizing channel $n$ times, (c) under generalized amplitude damping channel $n$ times.}
\end{center}
\end{figure}

Furthermore, if the first subsystem goes through the amplitude damping channel $n$ times, the output state $\rho_{adc}^{(n)}$ is $\rho_{adc}^{(n)}=E_0\otimes I \rho_{adc}^{(n-1)} E_0^\dagger\otimes I + E_1\otimes I \rho_{adc}^{(n-1)} E_1^\dagger\otimes I$, which can be rewritten as
$\rho_{adc}^{(n)}=\sum_{i_1,i_2,\cdots, i_n=0,1}  E_{i_1i_2\cdots i_n} \otimes I \rho^{ab} E_{i_1i_2\cdots i_n}^\dagger\otimes I$ with $E_{i_1i_2\cdots i_n}=E_{i_1}E_{i_2}\cdots E_{i_n}$. Using the properties of operators $E_0$ and $E_1$ in the amplitude damping channel,
$E_1^2=0$, $E_0E_1=E_1$ and $E_1E_0=\sqrt{1-p}E_1$, $\rho_{adc}^{(n)}$ is simplified to be $\rho_{adc}^{(n)}=E_0^n \otimes I \rho^{ab} (E_0^n)^\dagger\otimes I+\sum_{i=0}^{n-1} E_1 E_0^{n-i-1} \otimes I \rho^{ab} (E_1 E_0^{n-i-1})^\dagger \otimes I$, namely,
\begin{eqnarray}\label{left rho}
\rho_{adc}^{(n)}=\frac{1}{4}\left(
\begin{array}{cccc}
2-(1-c_{3})(1-p)^{n}& 0 & 0 & (c_{1}-c_{2})(1-p)^{\frac{n}{2}}\\
0& 2-(1+c_{3})(1-p)^{n} & (c_{1}+c_{2})(1-p)^{\frac{n}{2}}& 0\\
0& (c_{1}+c_{2})(1-p)^{\frac{n}{2}} & (1-c_{3})(1-p)^{n} & 0\\
(c_{1}-c_{2})(1-p)^{\frac{n}{2}}& 0 & 0 & (1+c_{3})(1-p)^{n}
\end{array}
\right). \nonumber
\end{eqnarray}

The eigenvalues of $\rho_{adc}^{(n)}$ are  $u_{0}^{(n)}=1/4(1-c_{3}(1-p)^{n}-\sqrt{(c_{1}+c_{2})^{2}(1-p)^{n}+(-1+(1-p)^{n})^{2}})$, $u_{1}^{(n)}=1/4(1-c_{3}(1-p)^{n}+\sqrt{(c_{1}+c_{2})^{2}(1-p)^{n}+(-1+(1-p)^{n})^{2}})$, $u_{2}^{(n)}=1/4(1+c_{3}(1-p)^{n}-\sqrt{(c_{1}-c_{2})^{2}(1-p)^{n}+(-1+(1-p)^{n})^{2}})$ and $u_{3}^{(n)}=1/4(1+c_{3}(1-p)^{n}+\sqrt{(c_{1}-c_{2})^{2}(1-p)^{n}+(-1+(1-p)^{n})^{2}})$.
Set $c_{1}=0.5, c_{2}=0.3, c_{3}=0.1, \epsilon^{A}=0.6$ and $\epsilon^{B}=0.3$. We have always $u^{(n)}_{0}<u^{(n)}_{2}<u^{(n)}_{3}<u^{(n)}_{1}$ for $n=2$, $n=3$, $n=4$, $n=10$ and $n=100$, see Fig. 4. (a)-(e). Using Eq. (\ref{zyf1}) we have
\begin{align}
C^{adc}_{123}(\rho^{(n)}_{adc},H_{AB})&=\sqrt{(c_{1}+c_{2})^{2}(1-p)^{n}
+(-1+(1-p)^{n})^{2}}(\epsilon^A+\epsilon^B)\nonumber\\
&~~~ +\sqrt{(c_{1}-c_{2})^{2}(1-p)^{n}+(-1+(1-p)^{n})^{2}}(\epsilon^A-\epsilon^B)\nonumber\\
&= 0.9\sqrt{0.64(1-p)^{n}+(-1+(1-p)^{n})^{2}}+0.3\sqrt{0.04(1-p)^{^{n}}+(-1+(1-p)^{n}}.\nonumber
\end{align}
One sees from Fig. 4. (f) that under amplitude damping channel n times the quantum battery capacity of Bell-diagonal states increases as $p$ and $n$ increases separately. It approaches to a constant for large $n$, that is, the frozen capacity appears.
\begin{figure}[h]
\begin{center}
\scalebox{1.0}{(a)}{\includegraphics[width=4.8cm]{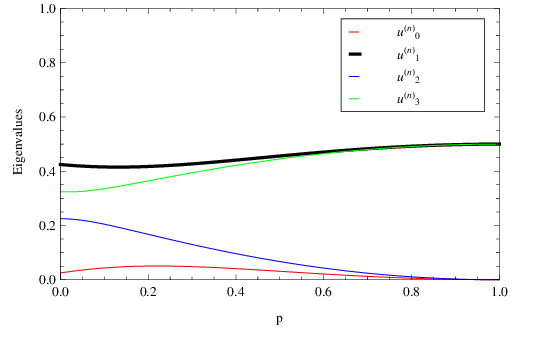}}
\scalebox{1.0}{(b)}{\includegraphics[width=4.8cm]{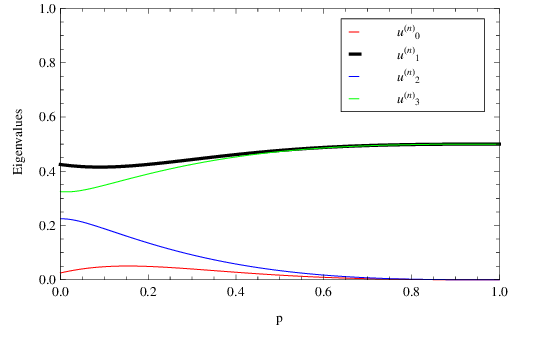}}
\scalebox{1.0}{(c)}{\includegraphics[width=4.8cm]{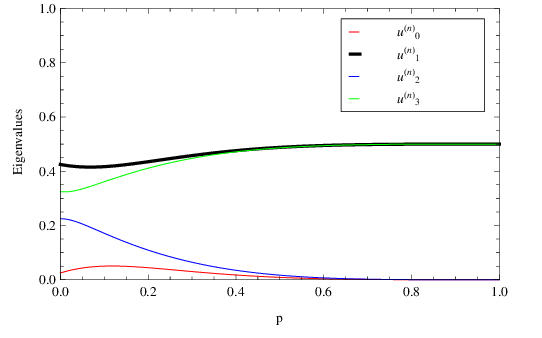}}
\scalebox{1.0}{(d)}{\includegraphics[width=4.8cm]{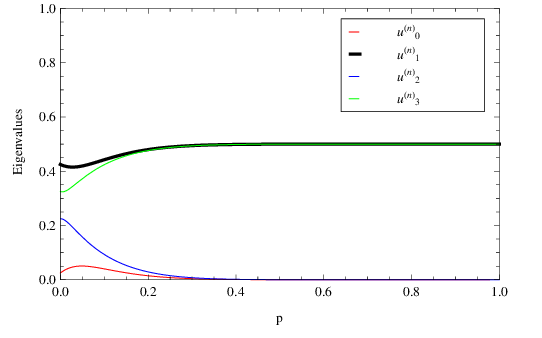}}
\scalebox{1.0}{(e)}{\includegraphics[width=4.8cm]{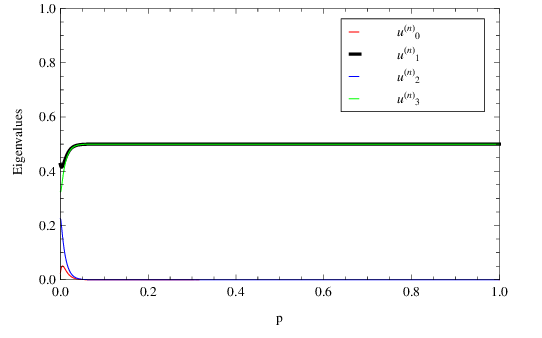}}
\scalebox{1.0}{(f)}{\includegraphics[width=4.8cm]{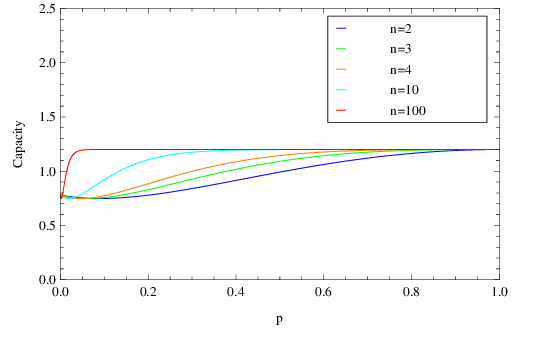}}
\caption{The eigenvalues of $\rho^{(n)}_{adc}$ as a function of $p$ for $c_{1}=0.5, c_{2}=0.3, c_{3}=0.1, \epsilon^{A}=0.6, \epsilon^{B}=0.3$ for (a) $n=2$, (b) $n=3$, (c) $n=4$, (d) $n=10$ and (e) $n=100$. (f) Quantum battery capacity evolution for Bell-diagonal states \{$c_{1}=0.5, c_{2}=0.3, c_{3}=0.1, \epsilon^{A}=0.6, \epsilon^{B}=0.3 $\} under amplitude damping channel n times as a function of $p$.}
\end{center}
\end{figure}

\section{Quantum battery capacity dynamics of Bell-diagonal states under Bi-side Markovian channels of the same type}

In this section, we focus on the dynamics of quantum battery capacity of a
two-qubit Bell-diagonal state undergoing two independent local Markovian channels of the
same type but with different decoherence rates, see the Kraus operators in Table~\ref{t4}.
\begin{table}
\begin{center}
\begin{tabular}{c c}
\hline \hline
 & $\textrm{Kraus operators}$                                         \\  \hline & \\
pf-pf   & $E_0^{(A)} = \sqrt{1-\frac{p}{2}} I^{A}\otimes I^{B},~~~ E_1^{(A)}=\sqrt{\frac{p}{2}}\sigma^A_3\otimes I^{B}$             \\
 & \\
  & $E_0^{(B)} =I^{A}\otimes\sqrt{1-\frac{q}{2}}I^{B},~~~  E_1^{(B)} =I^{A}\otimes\sqrt{\frac{q}{2}}\sigma^B_3$                 \\
 & \\
bf-bf   & $E_0^{(A)}=\sqrt{1-\frac{p}{2}}I^{A}\otimes I^{B},~~~ E_1^{(A)}=\sqrt{\frac{p}{2}}\sigma^A_1\otimes I^{B}$                \\
 & \\
 & $E_0^{(B)}=I^{A}\otimes\sqrt{1-\frac{q}{2}}I^{B},~~~ E_1^{(B)}=I^{A}\otimes\sqrt{\frac{q}{2}}\sigma^B_1$                 \\
 & \\

 bpf-bpf   & $E_0^{(A)}=\sqrt{1-\frac{p}{2}}I^{A}\otimes I^{B},~~~ E_1^{(A)}=\sqrt{\frac{p}{2}}\sigma^A_2\otimes I^{B}$                \\
 & \\
 & $E_0^{(B)}=I^{A}\otimes\sqrt{1-\frac{q}{2}}I^{B},~~~ E_1^{(B)}=I^{A}\otimes\sqrt{\frac{q}{2}}\sigma^B_2$                 \\
 & \\
\hline \hline
\end{tabular}
\caption[table 1]{Kraus operators for two independent local Markovian channels: two independent local phase-flip channels (pf-pf), two independent local bit-flip channels(bf-bf), two independent local bit-phase-flip channels (bpf-bpf), where $p=1-\text{exp}(-\gamma t)$, $q=1-\text{exp}(-\gamma't)$, and $\gamma$ and $\gamma'$ are the phase damping rates for the channels on the qubits $A$ and $B$, respectively.}
\label{t4}
\end{center}
\end{table}

Any Bell-diagonal state $\rho$ given by Eq. (\ref{bs}) evolves to another Bell-diagonal state Eq. (\ref{newstate}) under these channels. The corresponding coefficients are listed in the  Table \ref{t5}. As an example, let $c_{1}=0.5$, $c_{2}=0.3$, $c_{3}=0.1$, $\epsilon^{A}=0.6$ and $\epsilon^{B}=0.3$. From Eq. (\ref{123}) we have the dynamical behaviors of the quantum battery capacity for Bell-diagonal state under the bi-side same type Markovian
channel of bit flip, see Fig. 5. When $p$ and $q$ increase, the quantum battery capacity decreases and tends to constant. The quantum battery capacity for Bell-diagonal states under the bi-side channel of phase-flip and bit-flip is similar varied.
\begin{table}
\begin{center}
\begin{tabular}{c c c c}
\hline \hline
$\textrm{Channel}$ & $c^\prime_1$      & $c^\prime_2$     & $c^\prime_3$      \\ \hline
& & & \\
pf-pf                 &  $(1-p)(1-q)c_1$~~~~            & $(1-p)(1-q)c_2$    & $c_3$     \\
& & & \\
bf-bf                 &  $c_1$    &~~~ $(1-p)(1-q)c_2$ ~~~   & $(1-p)(1-q)c_3$             \\
& & & \\
bpf-bpf        ~~~        &  $(1-p)(1-q)c_1$    & $c_2$            & $(1-p)(1-q)c_3$     \\
& & & \\
 \hline \hline
\end{tabular}
\caption[table 2]{Correlation coefficients after the quantum channels: two independent local phase-flip channels (pf-pf), two independent local bit-flip channels (bf-bf), two independent local bit-phase-flip channels (bpf-bpf), where $p=1-\text{exp}(-\gamma t)$, $q=1-\text{exp}(-\gamma't)$, and $\gamma$ and $\gamma'$ are the phase damping rates for the channels on the qubits $A$ and $B$, respectively.}
\label{t5}
\end{center}
\end{table}
\begin{figure}[h]
\begin{center}
\scalebox{1.0}{\includegraphics[width=7cm]{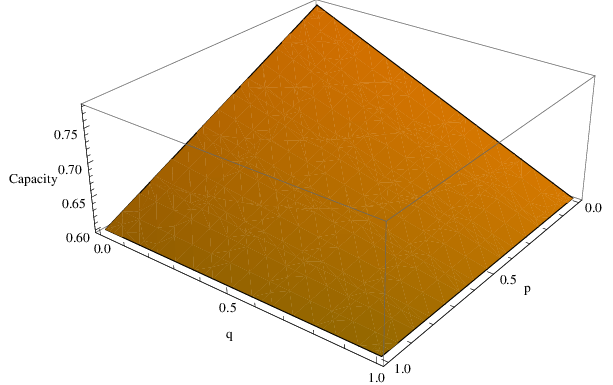}}
\caption{Quantum battery capacity for Bell-diagonal state \{$c_{1}=0.5, c_{2}=0.3, c_{3}=0.1, \epsilon^{A}=0.6, \epsilon^{B}=0.3$\} under bi-side Markovian channels of the bit flip channel.}.
\end{center}
\end{figure}

Moreover, if both subsystems go through the same type channel $n$ times, the Bell-diagonal state $\rho$ given by (\ref{bs}) also evolves to another Bell-diagonal state (\ref{newstate}). The corresponding coefficients are listed in Table \ref{t51}. As an example, let $c_{1}=0.5, c_{2}=0.3, c_{3}=0.1, \epsilon^{A}=0.6, \epsilon^{B}=0.3$. From Eq. (\ref{123}) we obtain the quantum battery capacity for Bell-diagonal state under bi-side same type Markovian channel $n$ times. In Fig. 6, we see that when $n$ becomes larger, the quantum battery capacity for Bell-diagonal state under bi-side same type Markovian channel decreases quickly as $p$ and $q$ increase. It also increases as $n$ increases. The quantum battery capacity for Bell-diagonal states under the bi-side channel of phase-flip and bit-flip $n$ times varies similarly.
\begin{table}
\begin{center}
\begin{tabular}{c c c c}
\hline \hline
$\textrm{Channel}$ & $c^\prime_1$      & $c^\prime_2$     & $c^\prime_3$      \\ \hline
& & & \\
$pf^{n}-pf^{n}$                 &  $(1-p)^{n}(1-q)^{n}c_1$            & $(1-p)^{n}(1-q)^{n}c_2$    & $c_3$     \\
& & & \\
$bf^{n}-bf^{n}$                &  $c_1$    & $(1-p)^{n}(1-q)^{n}c_2$    & $(1-p)^{n}(1-q)^{n}c_3$             \\
& & & \\
$bpf^{n}-bpf^{n}$                &  $(1-p)^{n}(1-q)^{n}c_1$    & $c_2$            & $(1-p)^{n}(1-q)^{n}c_3$     \\
& & & \\
 \hline \hline
\end{tabular}
\caption[table 2]{Correlation coefficients for $n$ times quantum operations: two independent local phase-flip channels ($pf^{n}-pf^{n}$), two independent local bit-flip channels ($bf^{n}-bf^{n}$), two independent local bit-phase-flip channels ($bpf^{n}-bpf^{n}$), where $p=1-\text{exp}(-\gamma t)$, $q=1-\text{exp}(-\gamma't)$, and $\gamma$ and $\gamma'$ are the phase damping rates for the channels on qubits $A$ and $B$, respectively.}
\label{t51}
\end{center}
\end{table}
\begin{figure}[!h]
\begin{center}
\scalebox{1.0}{(a)}{\includegraphics[width=6cm]{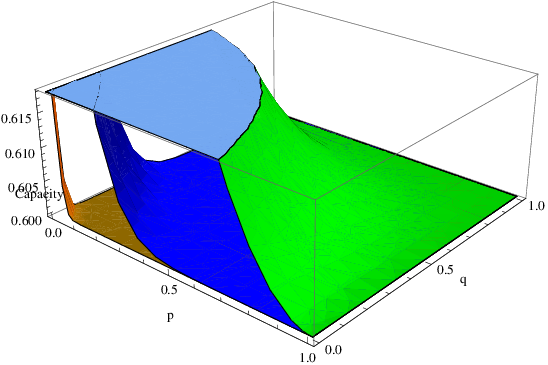}}
\caption{Quantum battery capacity for Bell-diagonal state \{$c_{1}=0.5, c_{2}=0.3, c_{3}=0.1, \epsilon^{A}=0.6, \epsilon^{B}=0.3$\} under bi-side Markovian channel of the bit flip channel $n$ times: $n=2$ [orange surface], $n=10$ [blue surface], $n=100$ [green surface].}
\end{center}
\end{figure}

\section{\bf summary}\label{IIII}

In this work, we have investigated the quantum battery capacity under the channels of bit flip, phase flip, bit-phase flip, depolarizing, amplitude damping and generalized amplitude damping on the first subsystem for Bell-diagonal states. In particular, we have shown that the quantum battery capacity of the Bell-diagonal states increases under amplitude damping channel. Moreover, the sudden death occurs under depolarizing channel.

We have also studied the quantum battery capacity evolution under Markovian channels on the first subsystem $n$ times. It has been shown that the quantum battery capacity of the Bell-diagonal states tends to a constant for large $n$ under bit flip channel and amplitude damping channel, namely, the frozen capacity shows up. Furthermore, we have studied the dynamics of quantum battery capacity of the Bell-diagonal states under two independent same type local Markovian channels of bit-flip, phase-flip and bit-phase-flip. Our results may highlight further investigations on the quantum battery capacity evolution under local quantum channels.

\bigskip
\noindent {\bf Acknowledgments}  This work was supported by the National Natural Science Foundation of China under grant No. 12371135, 12065021, 12075159 and 12171044, and the specific research fund of the Innovation Platform for Academicians of Hainan Province.

\end{document}